\documentclass{article}[12pt]
\usepackage{amsmath,amssymb}
\usepackage{graphicx}%
\usepackage{verbatim}
\begin{document}

\begin{center}
{\bf \Large Qualitative and Numerical Analysis of \\[6pt]
a Cosmological Model Based on an Asymmetric\\[6pt]
Scalar Doublet with Minimal Couplings.\\[6pt]
II. Numerical Modeling of Phase Trajectories} \\[12pt]
Yu. G. Ignat'ev and I. A. Kokh\\
N. I. Lobachevsky Institute of Mathematics and Mechanics of Kazan Federal University,\\
Kremleovskaya str., 35, Kazan, 420008, Russia.
\end{center}

%\maketitle

\begin{abstract}

With the help of our own software package DifEqTools, numerical modeling of the cosmological evolution of
a system consisting of an asymmetric scalar doublet of nonlinear, minimally interacting scalar fields,
a classical field and a phantom field, has been performed. Peculiarities of the behavior of the model near zeroenergy
hypersurfaces have been revealed.\\


\end{abstract}

\section{The basic equations of a cosmological model based on an asymmetric scalar doublet}

In [1] a qualitative analysis of a cosmological model based on an asymmetric scalar doublet of scalar fields --
a classical field ($\Phi$) and a phantom field ($\varphi$) minimally interacting with each other -- was performed. It was shown that
the corresponding 4-dimensional dynamical system can, depending on the parameters of the model, have 1, 3, or 9 stationary points lying in the $\{ \dot{\Phi }=0,\dot{\varphi }=0\}$ plane. In the given paper, we perform a numerical modeling of the cosmological evolution of a system consisting of an asymmetric scalar doublet and reveal the main unique types of
behavior of this system. The numerical modeling was realized in the applied mathematical software package Maple
with the help of the authors' program library DifEqTools [2], specially designed to model multidimensional nonlinear
dynamical systems.

In dimensionless variables, a closed system of ordinary differential equations describing the cosmological evolution of an asymmetric scalar doublet in the case of a spatially-flat Universe has the form [1]

\begin{equation} \label{_1_} 3\frac{a'^{2}}{a^{2} } =\left(\Phi'^{2} +e\Phi^{2} -\frac{\alpha_m}{2} \Phi^{4} \right)-\left(\varphi'^{2} -\varepsilon\mu^{2} \varphi^{2} +\frac{\beta_m }{2} \varphi^{4} \right)+\lambda_m ; \end{equation}

\begin{equation} \label{_2_} \Phi'' +3\frac{a'}{a} \Phi'+e \Phi -\alpha_m \Phi ^{3} =0; \end{equation}

\begin{equation} \label{_3_} \varphi''+3\frac{a'}{a} \varphi'-\varepsilon \mu^{2} \varphi +\beta_m \varphi ^{3} =0, \end{equation}
where $e,\varepsilon =\pm 1$;

\[\lambda _{m} \equiv \frac{\lambda }{m^{2} } ;\quad \alpha _{m} \equiv \frac{\alpha }{m^{2} };\quad \beta _{m} \equiv \frac{\beta }{m^{2} }; \quad \mu \equiv \frac{{\rm m}}{m} ;\]
$\alpha$ and $\beta $ are the self-action constants for the classical and the phantom fields, $m$  and ${\rm m}$ are the masses of these fields,
respectively, $\lambda$ is the cosmological constant, and the prime symbol denotes differentiation with respect to the
dimensionless time variable $\tau$, related to physical time $t$ by the formula
\[\tau =mt.\]
The time $\tau$ is measured in Compton units based on the classical scalar field. Here the following formulas involving the
scale factor $a(\tau)$ are valid:
\[\frac{a'}{a} \equiv L'=H_{m} \equiv \frac{H}{m}; \quad {\rm }\Omega =\frac{aa''}{a'^{2} } \equiv 1+\frac{H'_{m} }{H_{m} ^{2} } ,\]
where $H$ is the Hubble constant, $\Omega$ is the invariant cosmological acceleration, and $L=\ln a$. The contributions to
the total energy density of the classical field $(E_c)$ and the phantom field $(E_p)$ inside the parentheses in Eq.\eqref{_1_} are
respectively
\[E_{c} =m^{2} \left(\Phi '^{2} +e\Phi ^{2} -\frac{\alpha _{m} }{2} \Phi ^{4} \right); E _{p} =m^{2} \left(-\varphi '^{2} +\varepsilon \mu ^{2} \varphi ^{2} -\frac{\beta _{m} }{2} \varphi ^{4} \right);{\rm \; \; }  \]
\[ p_{c} =m^{2} \left(\Phi '^{2} -e\Phi ^{2} +\frac{\alpha _{m} }{2} \Phi ^{4} \right);{\rm }p_{p} =m^{2} \left(-\varphi '^{2} -\varepsilon \mu ^{2} \varphi ^{2} +\frac{\beta _{m} }{2} \varphi ^{4} \right). \]
Introducing a potential energy of the classical field $(V(\Phi ))$ and the phantom field $(v(\varphi ))$ in line with [1]
\begin{equation} \label{_4_}
V(\Phi )=-\frac{\alpha }{4} \left(\Phi ^{2} -e\frac{m^{2} }{\alpha } \right)^{2} ,{\rm \; \; \; }v(\varphi )=-\frac{\beta }{4} \left(\varphi ^{2} -\varepsilon \frac{{\rm m}^{2} }{\beta } \right)^{2} ,
\end{equation}
we write down an expression for the reduced ($E_{m} =\varepsilon /m^{2} $) effective energy density
\begin{equation} \label{_5_}
E_{m} =E_{c} +E_{p} +\Lambda _{m} =\left(\frac{Z^{2} }{2} -V(\Phi )\right)-\left(\frac{z^{2} }{2} -v(\varphi )\right)+\Lambda _{m} ,
\end{equation}
where
\begin{equation} \label{_6_}
\Lambda _{m} =\lambda _{m} -\frac{1}{2\alpha _{m} } -\frac{\mu ^{2} }{2\beta _{m} }
\end{equation}
and we have introduced  \textit{total energies} for the classical and the phantom field:
\begin{equation} \label{_7_}
E_{c} =\frac{Z^{2} }{2} -V(\Phi ),{\rm \; \; \; \; }E_{p} =-\left(\frac{z^{2} }{2} -v(\varphi )\right),{\rm \; \; \; \; }E_{m} =E_{c} +E_{p} +\Lambda _{m} .
\end{equation}
Thus, the Einstein equation (Eq. \eqref{_1_}) can be rewritten in dimensionless form:
\begin{equation} \label{_8_}
\frac{a'^{2} }{a^{2} } \equiv H_{m}^{2} =\frac{1}{3} E_{m} .
\end{equation}
In dimensionless variables, we can write the closed normal system of ordinary differential equations describing the cosmological evolution of the asymmetric scalar doublet in the case of a spatially flat Universe as [1]
\begin{equation} \label{_9_}
\begin{array}{c} {\Phi '=Z,} \\[12pt] {Z'=-\sqrt{3} Z\sqrt{\left(Z^{2} +e\Phi ^{2} -\displaystyle\frac{\alpha _{m} }{2} \Phi ^{4} \right)-\left(z^{2} -\varepsilon \mu ^{2} \varphi ^{2} +\displaystyle\frac{\beta _{m} }{2} \varphi ^{4} \right)+\lambda _{m} } -e\Phi +\alpha _{m} \Phi ^{3} ,} \\[12pt] {\varphi '=z,} \\[12pt] {z'=-\sqrt{3} z\sqrt{\left(Z^{2} +e\Phi ^{2} -\displaystyle\frac{\alpha _{m} }{2} \Phi ^{4} \right)-\left(z^{2} -\varepsilon \mu ^{2} \varphi ^{2} +\displaystyle\frac{\beta _{m} }{2} \varphi ^{4} \right)+\lambda _{m} } +\varepsilon \mu ^{2} \varphi -\beta _{m} \varphi ^{3} .} \end{array}
\end{equation}
In these variables, the Einstein equation takes the following form:
\begin{equation} \label{_10_}
H'_{m} {}^{2} =\frac{1}{3} \left[\left(Z^{2} +e\Phi ^{2} -\frac{\alpha _{m} }{2} \Phi ^{4} \right)-\left(z^{2} -\varepsilon \mu ^{2} \varphi ^{2} +\frac{\beta _{m} }{2} \varphi ^{4} \right)+\lambda _{m} \right].
\end{equation}

\section{Real-valued regions and motion near energy hypersurfaces }

\noindent As was noted in [1], a unique peculiarity of the considered system is the variation of the topology of the phase space as a consequence of the appearance in it of regions in which motion is impossible.  These regions are distinguished by the condition of negativity of the effective total energy (Eq. (5)), and the regions accessible to the phase trajectories are defined by the condition
\begin{equation} \label{_11_}
E_{m} =E_{c} +E_{p} +\Lambda _{m} \ge 0.
\end{equation}
Hypersurfaces of zero effective energy $S_{3}^{0} \subset {\rm R}_{4} $, separating phase space into regions of accessible and forbidden values of the dynamic variables, are described by the equations
\begin{equation} \label{_12_}
E_{m} =0\Rightarrow \frac{Z^{2} }{2} -V(\Phi )-\left(\frac{z^{2} }{2} -v(\varphi )\right)+\Lambda _{m} =0.
\end{equation}
Note that as a consequence of definition \eqref{_6_}, the renormalized value of the cosmological constant $\Lambda _{m} $ can also take, generally speaking, negative values.    Field equations \eqref{_2_} and \eqref{_3_} have the form of equations of free oscillations in the field of a fourth-order potential
\begin{equation} \label{_13_}
\ddot{x}+k\dot{x}+\frac{\partial V}{\partial x} =0
\end{equation}
with a \textit{nonlinear friction coefficient}
\begin{equation} \label{_14_}
k=\sqrt{3E_{m} } \equiv \sqrt{3(E_{c} (\Phi ,Z)+E_{p} (\varphi ,z)+\Lambda _{m} )} .
\end{equation}
According to the theory of oscillations, the corresponding dynamical system should, by losing total energy as a consequence of a dissipative process corresponding to the friction force in Eq. \eqref{_13_}, fall over the course of time into a potential energy minimum in the presence of the latter and, in the case of its absence, to unlimited rolling \textit{downward}.  The specifics of our problem consist, first of all, in the fundamental dependence of the friction coefficient on the total energy of the system and, secondly, in the nonlinear coupling of the subsystems through the \textit{friction coefficient} and, thirdly, in the factor of negativity of the kinetic energy of the phantom component.

Let us consider motion along the surface of zero effective energy (Eq. (12)).  Equations \eqref{_9_} take the following form on this surface:
\begin{equation} \label{_15_}
\begin{array}{c} {\Phi '=Z,} \\[12pt] {Z'=-e\Phi +\alpha _{m} \Phi ^{3} ,} \\[12pt] {\varphi '=z,} \\[12pt] {z'=\varepsilon \mu ^{2} \varphi -\beta _{m} \varphi ^{3} .} \end{array}
\end{equation}
The right-hand sides of the second and fourth of Eqs. \eqref{_15_} are derivatives with respect to the scalar fields of the corresponding potential functions.  Indeed, differentiating relations \eqref{_4_}, we obtain
\[\frac{dV}{d\Phi } =e\Phi -\alpha _{m} \Phi ^{3} ,{\rm \; \; \; \; }\frac{dv}{d\varphi } =\varepsilon \mu ^{2} \varphi -\beta _{m} \varphi ^{3} .\]
Thus, multiplying the second and fourth of Eqs. \eqref{_15_} by \textit{Z} and \textit{z}, respectively, we obtain total energy integrals for motion along the hypersurface $S_{3}^{0} $:
\begin{equation} \label{_16_}
\begin{array}{c}\displaystyle {\frac{Z^{2} }{2} -V(\Phi )=\frac{Z^{2} }{2} -\frac{\alpha _{m} }{4} \left(\Phi ^{2} -\frac{e}{\alpha _{m} } \right)^{2} =E_{c} =const,} \\[12pt] \displaystyle {\frac{z^{2} }{2} -v(\varphi )=\frac{z^{2} }{2} +\frac{\beta _{m} }{4} \left(\varphi ^{2} -\frac{\varepsilon \mu }{\beta _{m} } \right)^{2} =-E_{p} =const.} \end{array}
\end{equation}
This fact at once tells us that we are dealing here with free oscillations. As a consequence of Eqs. \eqref{_12_}, the total energy integrals on this trajectory should be related as follows:
\begin{equation} \label{_17_}
E_{c} +E_{p} +\Lambda _{m} =0.
\end{equation}

As a result, we can assert that the phase trajectories on the zero energy surface, described by Eqs. \eqref{_15_} with total energy integrals \eqref{_16_} and \eqref{_17_}, are the exact solution of the total equations of motion (Eqs. (9)).  To obtain a concrete phase trajectory, it is necessary to assign one of the constants ($E_{c} $ or $E_{p} $) and find the second constant from relation \eqref{_17_}.  Thus, the desired phase trajectories are cross sections of the zero effective energy surface (Eq. (12)).  It is also possible to obtain explicit solutions of the field equations on the effective energy surface (Eq. (12)) by integrating Eqs. \eqref{_16_}:
\[\begin{array}{c} \displaystyle{\Phi =\Phi _{0} \pm \int\limits_{0}^{\tau }\sqrt{2E_{c} +\frac{\alpha _{m} }{2} \left(\Phi ^{2} -\frac{e}{\alpha _{m} } \right)^{2} }  ,} \\[12pt] \displaystyle{\varphi =\varphi _{0} \pm \int\limits_{0}^{\tau }\sqrt{2E_{p} -\frac{\beta _{m} }{2} \left(\varphi ^{2} -\frac{\varepsilon \mu }{\beta _{m} } \right)^{2} }  .} \end{array}\]

The result of integrating these two equations is expressed with the help of elliptical functions. It is to be expected that in the case in which we find attractive centers inside forbidden regions, the phase trajectories will adhere to the zero effective energy surfaces whereas in the case in which we find saddle points in these regions, they will be repulsed from the zero effective energy surface.

\section{ Numerical modeling of a dynamical system   }

We present numerical integration results demonstrating the indicated peculiarities.  Figures 1 and 2 show phase trajectories of a phantom field (Fig. 1) and a classical field (Fig. 2).  The dot-dash curves depict the projections of the zero effective energy surface.

 \begin{figure}[h!]
 \centerline{\includegraphics[width=8.5cm]{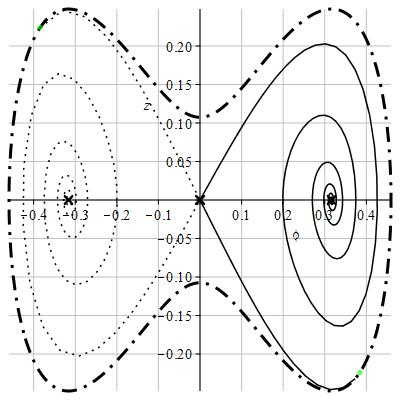}\label{Fig2}} \caption{Phase trajectories of the phantom field near a zero energy surface (Eqs. (12)):  $\Phi (0)=\pm 0.05,{\rm \; }Z(0)=0,\; \; \varphi (0)=\pm 0.000001,\; \, {\rm and}\, \; z(0)=0$.
 }
\end{figure}
 \begin{figure}[h!]
 \centerline{\includegraphics[width=8.5cm]{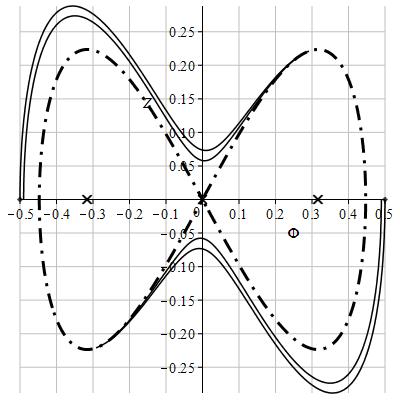}\label{Fig2}} \caption{Phase trajectories of the classical field near a zero energy surface (Eqs. (12)): $\Phi (0)=\pm 0.5;\; 0.49;$ $Z(0)=0,$ $\varphi (0)=\pm 0.00001,$ and $z(0)=0$.
 }
\end{figure}

The phase trajectories presented here were obtained with the help of the special applied software package DifEqTools [2] for the following values of the parameters of the doublet: $\alpha _{m} =10,\, \, \beta _{m} =10$, ${\rm \; }e=1,{\rm \; }\varepsilon =1,{\rm \; }\mu =1,{\rm \; }\lambda =0$ for the phase trajectory in Fig. 1 and  $\alpha _{m} =-10,\, \, \beta _{m} =10,{\rm \; }e=-1,{\rm \; }\varepsilon =1,{\rm \; }\mu =1$, $\lambda =0$ for the phase trajectory in Fig. 2.  The crosses in these figures mark the singular points of the system, and the origin of the coordinate system is a saddle point.  As can be seen, the phantom field begins its history near a zero energy surface, and then, after being repulsed from the saddle point, it approaches, along a spiral path, the attractive centers located for the phantom field inside the allowed region.  The classical field, in contrast, begins its history far from the zero energy surface, the interior of which is forbidden for the classical field, and then gradually lays up upon this surface. Figures 3 and 4 show graphs of the evolution of the effective energy (solid curve) of the system for different initial conditions and parameters of the doublet.  The open circles represent values of the energy of the phantom field, and the filled circles represent the energy of the classical field. In the first graph, the effective energy tends to the constant value 0.05 in the infinite future, and in the second graph, to zero.

 \begin{figure}
 \centerline{\includegraphics[width=8.5cm]{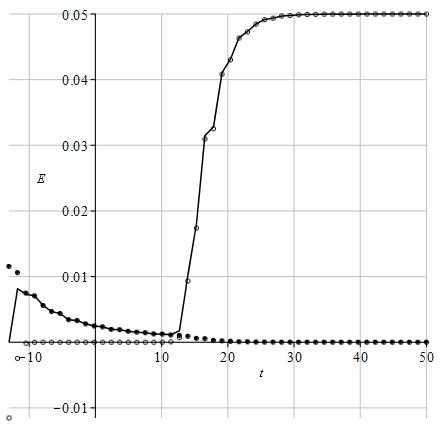}\label{Fig2}} \caption{Evolution of the effective energy of the system: $\alpha _{m} =10,\, \, \, \beta _{m} =10,{\rm \; }e=1,{\rm \; }\varepsilon =1,{\rm \; }\mu =1,{\rm \; }\lambda =0$, $\Phi (0)=0.05,{\rm \; }Z(0)=0,\; \; \varphi (0)=\pm 0.000001,\; \; {\rm and}\; {\kern 1pt} z(0)=0$.
 }
\end{figure}

 \begin{figure}
 \centerline{\includegraphics[width=8.5cm]{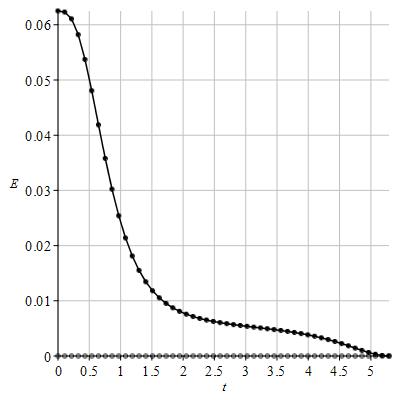}\label{Fig2}} \caption{Evolution of the effective energy of the system: $\alpha _{m} =-10,\, \, \, \beta _{m} =10,{\rm \; }e=-1,{\rm \; }\varepsilon =1,{\rm \; }\mu =1,{\rm \; }\lambda =0$, $\Phi (0)=0.5,{\rm \; }Z(0)=0,\; \; \varphi (0)=\pm 0.000001,\; \; {\rm and}\; \, z(0)=0$.
 }
\end{figure}

In conclusion we can state with assurance that we have confirmed and refined the main conclusions of [1].  Along with that, the main asymptotic properties of the cosmological model based on the asymmetric doublet $\{ \Phi ,\varphi \} $ have become physically more comprehensible. In particular, in such a model the phantom field can start practically from zero energy and then approach after an abrupt bounce some constant value, while the classical field can, in contrast, start from some initial energy but then in the course of its evolution wind up with this energy diminished to zero.  Thus we see very important features of the model of a scalar doublet which need to be examined more closely to create an adequate cosmological model.  In future work, it is our intention to carry out a more comprehensive study of cosmological models based on an asymmetric scalar doublet.

This work was performed in accordance with the Russian Government Program of Competitive Growth of Kazan Federal University.

%\end{comment}
\end{document}